# Title: Giant lattice softening at a Lifshitz transition in $Sr_2RuO_4$

**Authors:** Hilary M.L. Noad[1]*, Kousuke Ishida[1]†, You-Sheng Li[1], Elena Gati[1], Veronika C. Stangier[2], Naoki Kikugawa[3], Dmitry A. Sokolov[1], Michael Nicklas[1], Bongjae Kim[4], Igor I. Mazin[5,6], Markus Garst[7,8], Jörg Schmalian[2,8]*, Andrew P. Mackenzie[1,9]*, Clifford W. Hicks[1,10]*

**Affiliations:**

[1]Max Planck Institute for Chemical Physics of Solids; 01187 Dresden, Germany.

[2]Institut für Theorie der Kondensierten Materie, Karlsruher Institut für Technologie; 76131 Karlsruhe, Germany.

[3]National Institute for Materials Science; Tsukuba, Ibaraki 305-0003, Japan.

[4]Department of Physics, Kunsan National University; Gunsan, 54150, Korea.

[5]Department of Physics and Astronomy, George Mason University; Fairfax, VA 22030, USA.

[6]Quantum Science and Engineering Center, George Mason University; Fairfax, VA 22030, USA.

[7]Institut für Theoretische Festkörperphysik, Karlsruher Institut für Technologie; 76131 Karlsruhe, Germany.

[8]Institut für QuantenMaterialien und Technologien, Karlsruher Institut für Technologie; 76131 Karlsruhe, Germany.

[9]Scottish Universities Physics Alliance, School of Physics and Astronomy, University of St Andrews; St Andrews KY16 9SS, UK.

[10]School of Physics and Astronomy, University of Birmingham; Birmingham B15 2TT, UK.

*Corresponding authors. Email: Hilary.Noad@cpfs.mpg.de, joerg.schmalian@kit.edu, Andy.Mackenzie@cpfs.mpg.de, C.Hicks.1@bham.ac.uk

†Present address: Institute for Materials Research, Tohoku University; Sendai 980-8577, Japan.

**Abstract:** The interplay of electronic and structural degrees of freedom in solids is a topic of intense research. Experience and intuition suggest that structural changes drive conduction electron behavior, because the large number of valence electrons dominate the structural properties. As part of a seminal paper written over sixty years ago, Lifshitz discussed an alternative possibility: lattice softening driven by conduction electrons at topological Fermi surface transitions. The effect he predicted, however, was small, and has not been convincingly observed. Using measurements of the stress-strain relationship in the ultra-clean metal $Sr_2RuO_4$, we reveal a huge softening of the Young's modulus at a Lifshitz transition of a two-dimensional Fermi surface, and show that it is indeed entirely driven by the conduction electrons of the relevant energy band.

**One-Sentence Summary:** Conduction electrons drive a nonlinear elastic response in the ultra-clean, two-dimensional metal $Sr_2RuO_4$.

**Main Text:** The coupling between elastic and electronic degrees of freedom is crucial to determining the phase diagrams of correlated electron systems, for example those displaying electronic nematicity, in which conduction electrons develop anisotropies (*1*). However, there is always a 'chicken and egg' question – does the lattice drive, or respond to, the conduction electron physics (*2, 3*)? Here, we approach the entanglement of electronic and structural degrees of freedom using a different method to those most commonly employed. We study the stress-strain relationship of the quasi-two-dimensional correlated metal $Sr_2RuO_4$ as it is tuned through a saddle point Lifshitz transition (*4–6*) in which the Fermi surface topology changes and the Fermi level crosses a Van Hove singularity (VHS) (*7*). By combining direct stress-strain measurements with experimentally-determined entropy data across the same transition, we demonstrate the existence of an unexpectedly large softening of the lattice, driven entirely by conduction electrons. The possibility of such effects was discussed theoretically by Lifshitz himself over sixty years ago, but imagined to be extremely small (*4*). Here it is large and, in principle, singular, i.e. capable of introducing a lattice instability in the $T \to 0$ limit if not cut off by a phase transition to some other form of order. We discuss our results in the framework of quantum critical elasticity, and show that superconductivity is a natural way of cutting off quantum critical lattice softening.

The material platform for our experiments, $Sr_2RuO_4$, has attracted considerable attention both as an unconventional superconductor (*8–12*) and a benchmark two-dimensional Fermi liquid (*13, 14*). It is one of the cleanest correlated electron materials known. The best single crystals have residual resistivities of ≈ 50 nΩ-cm, corresponding to electron mean free paths of 2 μm or more (*15*). Its Fermi surface consists of three cylinders, commonly referred to as α, β and γ (*14*). Electronic correlations lead to a substantial mass renormalization over the values predicted in independent-electron band structure calculations, meaning that under ambient conditions the Fermi energy $E_F$ of the γ band sits only 14 meV below a saddle point Van Hove singularity at the M point of the tetragonal Brillouin zone (*16*). It is possible to tune this Van Hove singularity through $E_F$ by applying 0.7 ± 0.1 GPa along the [100] crystalline direction (*17, 18*), profoundly affecting the electronic properties. For example, the superconducting transition temperature is enhanced by a factor of 2.4 from its ambient-pressure value to 3.5 K (*19*), and the temperature dependence of the resistivity undergoes a large change from the conventional Fermi liquid $T^2$ dependence seen at higher and lower stress (*5*).

To probe the consequences of this Lifshitz transition on the lattice stiffness, we used bespoke apparatus in which both uniaxial stress $\sigma$ and strain $\varepsilon$ can be monitored simultaneously (*20*), allowing measurement of the Young's modulus (Fig. S1). To maximize the quantitative accuracy of the data, samples are milled into a necked shape, as shown in the inset of Fig. 1A, using a Xe plasma source focused ion beam. The end tabs are then embedded in epoxy, which acts as a conformal layer through which large forces can be transmitted to the brittle samples. The necking creates a rapid crossover between low- and high-stress regions of the sample, which is important for resolving fine features in the stress-strain relationship (*21*). Full details on how we extract the Young's modulus and strain of the sample, including an examination of possible systematic errors that could affect the analysis, can be found in the Supplementary Materials (*22*) and Figs. S2-S5.

Our core result is shown in Fig. 1A: the differential Young's modulus $E_x = d\sigma_{xx}/d\varepsilon_{xx}$ as a function of strain at 4 K. To demonstrate repeatability, data from three samples are shown. Samples 2 and 3 had a higher aspect ratio (see Supplementary Information), and are therefore expected to yield more accurate data. Each sample was in a different stress cell. The force calibrations of the cells were refined using the known Lifshitz stress of -0.7 GPa (*18*), where negative values denote compression. At the Lifshitz transition strain of $\varepsilon_{VHS} = (0.45 \pm 0.05)\%$, $E_x$ is seen to drop to around 146 GPa, and then beyond the transition to increase to around 200 GPa. In other words, contrary to our naïve expectation, the dip in $E_x$ is not a small effect; the softening of the lattice at the Lifshitz transition is between 10% and 15% depending on the definition used for the background value. Under tensile strain, $E_x$ again decreases quite rapidly. There is expected to be a Lifshitz transition under tensile strain equivalent to that under compressive strain, and this decrease in $E_x$ is most likely due to approach to that transition. A second striking aspect of the lattice softening at the Lifshitz strain is its strong temperature dependence, shown in Fig. 1B. At 40 K the dip is barely resolved. As the temperature is lowered, it sharpens and deepens, with substantial change observable even between 5 K and 4 K.

Elastic constants are second derivatives of the free energy: $C = \frac{\partial^2 F}{\partial \varepsilon^2}$. $F = U - TS$, where $U$ is the internal energy and $S$ the entropy. The intuitive expectation might be that the valence band contributions to $U$ completely dominate $F$, and so also the elastic moduli. However, the generally valid expression for the free energy $F(T,\varepsilon) = E_0(\varepsilon) - \int_0^T S(T',\varepsilon)\,dT'$, where $E_0$ is the ground state energy, reveals nicely that any temperature dependence of elastic constants derives from changes in the entropy, even if $E_0(\varepsilon)$ may dominate the elastic moduli. Our Young's modulus data have a strong temperature dependence (Fig. 1B), so a link between the Young's modulus and a strain-dependent entropy is expected. In Fig. 1C we compare the entropy obtained from a recent study of the elastocaloric effect on Sr$_2$RuO$_4$ (*23*) at 4 K with the Young's modulus data at the same temperature. The strong correlation between the two leads to the conclusion that, remarkably, the key physics we are observing is driven by a conduction band. At this low temperature, the phonon contribution to the entropy in Sr$_2$RuO$_4$ is negligible (*24*), and the valence band contribution even more so. All that will be observed is conduction band entropy, due to the density of states at the Fermi level. A quantitative analysis comparing the entropy and Young's modulus data can be found in the Supplementary Materials.

In order to understand the observed behavior quantitatively, we make use of a two-dimensional model for the Landau quasiparticles of the γ band (*23*) whose parameters are all constrained by other observations on Sr$_2$RuO$_4$ (*25*). This model, which is summarized in the Supplementary Materials, yields a contribution $F_\gamma(T, \varepsilon_{xx}, \varepsilon_{yy}, \varepsilon_{zz})$ to the electronic free energy as function of temperature and the three uniaxial strain values. The crucial ingredient of the theory is a symmetry adapted deformation potential $\alpha \sim t^{-1} \partial t / \partial \varepsilon$ with tight-binding hopping parameters $t$. (Notice that, while we mostly refer to the uniaxial strain along the *x*-axis as $\varepsilon$, we briefly write out the axis labels explicitly to account for the correct Poisson effects that enter any measurement of the Young's modulus.) The total free energy is then given as $F = F_0 + F_\gamma$ where we determine $F_0$ such that our model reproduces the correct elastic constants for the unstrained samples at the reference temperature of $T = 4$ K. We include the applied uniaxial stress via $F \to F - \varepsilon_{xx}\sigma_{xx}$, which yields the equation of state $\sigma_{xx} = \partial F / \partial \varepsilon_{xx}$. In linear elasticity, strain orthogonal to the applied stress is accounted for by Poisson ratios, such as $\varepsilon_{\kappa\kappa} = -\nu_{x\kappa}\varepsilon_{xx}$ with $\kappa = y, z$. For the rather large stress values applied here and given the subtle behavior near the Lifshitz transition, we must, however, allow for non-linear relations $\varepsilon_{\kappa\kappa} = \varepsilon_{\kappa\kappa}(\varepsilon_{xx})$. Those follow from $\partial F / \partial \varepsilon_{yy} = \partial F / \partial \varepsilon_{zz} = 0$. This finally yields the differential Young's modulus

$$E_x = \frac{d\sigma_{xx}}{d\varepsilon_{xx}} = C_{11} + C_{12}\frac{\partial \varepsilon_{yy}(\varepsilon_{xx})}{\partial \varepsilon_{xx}} + C_{13}\frac{\partial \varepsilon_{zz}(\varepsilon_{xx})}{\partial \varepsilon_{xx}} \tag{1},$$

with the usual definition of the elastic tensor. The elastic constants $C_{ij} = C_{0,ij} - \frac{2}{k_B T} \int \frac{d^3k}{(2\pi)^3} f_k (1-f_k) \frac{\partial E_k}{\partial \varepsilon_{ii}} \frac{\partial E_k}{\partial \varepsilon_{jj}}$ consist of a background contribution $C_{0,ij}$ and the part due to the $\gamma$-band with strain-dependent $E_k$ and Fermi function $f_k$. Within linear elasticity, the derivatives are the strain-independent Poisson ratios and $1/E_x$ is the 11 element of the inverse elastic tensor. We show in the Supplementary Materials that near the Lifshitz point $-\varepsilon_{yy}/\varepsilon_{xx}$ and the differential Poisson ratio $-\partial \varepsilon_{yy}/\partial \varepsilon_{xx}$ differ and have a pronounced strain dependence (Fig. S6). Once these non-linear Poisson's ratios are included, our model makes the predictions shown in Figs. 2A and B. The agreement with the experimental data shown in Fig. 1 is very good (see also Fig. S7). The temperature dependence of the dip in the Young's modulus and the relationship between the Young's modulus and entropy at 4 K are reproduced so well that they provide powerful evidence that the model correctly captures the key physics of the observations. Qualitatively, the softening at the Van Hove point is a consequence of the fact that Young's modulus is the sum of a presumed weakly temperature- and strain-dependent background contribution, $E_x^{(0)}$, and a singular conduction electron contribution from the $\gamma$ band:

$$E_x \cong E_x^{(0)} - A \log \frac{1}{(T/T_0)^2 + (\varepsilon_{xx} - \varepsilon_{VHS})^2} \qquad (2),$$

where $A$ is a positive constant and $T_0$ a constant of order the bandwidth. The logarithmic $T$- and strain dependence is due to the fact that the electronic contribution to the elastic constants is proportional to the density of states [see the $f_k(1-f_k)$ term in the $\gamma$-band contribution to $C_{ij}$], which diverges logarithmically at a Van Hove singularity in two dimensions. All electronic contributions to the $C_{ij}$ and the differential Poisson's ratios also show this singular, logarithmic dependence. The sign of the coefficient $A > 0$ reflects that the $\gamma$-band will always cause a softening of the important diagonal elements $C_{ii}$ of the elastic tensor. The magnitude of $A$ is determined by a combination of band renormalization factors and the deformation potential: $A \propto \alpha^2$.

A further prediction of the model is that the temperature dependence of the Young's modulus at the Lifshitz pressure has negative curvature. As seen in Fig. 2C, this is also observed. This negative curvature, plus the strong link between elastic and electronic degrees of freedom that our data have established, imply that the Young's modulus is related to an electronic susceptibility. The logarithmic softening cannot continue down to the lowest temperatures. This might be a first-order structural transition – initially proposed by Lifshitz – or the formation of some electronic order that prevents a mechanical instability. One way in which the logarithm will be cut off in a quasi-two-dimensional material is by coherent three-dimensional effects due to interlayer hopping. However, in the highly two-dimensional $\gamma$ band of $Sr_2RuO_4$, the scale for such processes is < 3 K (*8*). Whether the onset of superconductivity at $T_c = 3.5$ K is related to the mechanical stability of $Sr_2RuO_4$ is therefore an exciting open question.

In this context it was interesting to extend our Young's modulus measurements to its superconducting state, in which a small gap is opened at the Fermi energy. As shown in Fig. 3A and B, the strong normal-state softening of the lattice is indeed cut off by the onset of the superconductivity, with the lattice hardening again slightly below the superconducting $T_c$, an effect that is most pronounced at the Van Hove strain.

A final aspect of our measurements is shown in Fig. 3C, in which we present $E_x$ data over a wider range of temperatures, at zero strain, an intermediate strain, and the Van Hove strain. At zero strain, there is a broad minimum in the Young's modulus at $T \sim 40$ K, reflecting the fact that, unusually, $Sr_2RuO_4$ softens along the [100] crystalline direction as the temperature is decreased from room temperature (*26*). We show in Fig. S8 that this feature is reproduced in our model, demonstrating that it, too, is a consequence of the conduction electrons in the $\gamma$ sheet.

Although the model we have used to analyze our Sr$_2$RuO$_4$ data is specific to the Lifshitz transition in two dimensions, it can be viewed from a more general perspective that highlights the significance of low-temperature entanglement between electronic and elastic degrees of freedom and emphasizes the close connection between elastic response and entropy. Consider a quantum critical point (QCP) that can be crossed by varying some combination of elements of the strain tensor. Under the assumption that hyperscaling holds near any such a strain-tuned QCP, for the singular contribution to the entropy,

$$S(T,\varepsilon) = T^{d/z}\phi\left(\frac{\varepsilon - \varepsilon_c}{T^{1/\nu z}}\right) \qquad (3),$$

where $z$ and $\nu$ are the dynamical and correlation length exponents, respectively, $d$ is the dimension of space, $\phi$ is a universal scaling function, and $\varepsilon_c$ is the critical strain. The most dramatic consequence of Eq. 3 follows as one integrates the entropy with respect to temperature to obtain the free energy and then determines the elastic constant. Right at the QCP, where $|\varepsilon - \varepsilon_c|^{\nu z} \ll T$, it follows that

$$C = C_0 - \frac{\nu z}{\nu(z+d)-2}\phi''(0)\, T^{\frac{\nu(z+d)-2}{\nu z}} \qquad (4),$$

with $\phi''$ the second derivative of $\phi$. Here, $C_0$ is a temperature-independent background contribution to the elastic constant that enters as an integration constant for the free energy. Since the entropy is maximal at the QCP, $\phi'' < 0$. If $\nu(d+z) > 2$ the universal temperature correction is small and positive and the system is mechanically stable. On the other hand, the system must undergo an instability, defined by a vanishing elastic constant at a nonzero temperature, if the quantum Harris criterion $\nu(d+z) < 2$ (*27, 28*) is fulfilled. Then, the above scaling theory of a 'naked' QCP ceases to be valid; the system either crosses over to a new critical regime where strain becomes a genuine dynamical quantum critical mode or undergoes a phase transition to a new state of matter. As discussed in the Supplementary Materials and Figs. S9-S10, the Lifshitz point of this paper corresponds to $d = z = \nu^{-1} = 2$, placing us at the boundary of the quantum Harris criterion, resulting in a logarithmic temperature dependence, also leading to a mechanical instability. If one takes the limit $\nu(d+z) \to 2$, appropriate for the two-dimensional Lifshitz transition, the exponent in the temperature-dependent term in Eq. (4) does not simply vanish, as the prefactor $\frac{1}{\nu(z+d)-2}$ diverges at the same time. Instead, one recovers the logarithmic behavior of the Lifshitz transition.

Returning to the discussion of the experimental findings, the results presented in this paper constitute conclusive observation of a lattice softening driven by conduction electrons, the possibility of which was foreseen by Lifshitz over sixty years ago (*4*). However, he considered hydrostatic pressure and transitions in materials with three-dimensional electronic structure, in which case the relative change in bulk modulus would be $\sim 10^{-4}$, three orders of magnitude smaller than what we observe (See Supplementary Materials and Figs. S11-S12). Partly for this reason, previous searches using hydrostatic pressure were unable to unambiguously resolve the predicted effect (*29–33*). Why, then, does it give such a prominent experimental signature in Sr$_2$RuO$_4$? Firstly, we have worked with uniaxial rather than hydrostatic pressure. Also, Sr$_2$RuO$_4$ is an extremely clean material for which the relevant band is strongly two-dimensional, preventing the logarithmic term in Eq. 2 from being washed out by three dimensional effects or disorder broadening. It is tempting on first inspection to assume that this logarithm makes the dominant contribution to the size of our signal, but for measurements performed at a few K, the size of the prefactor $A$ actually plays the crucial role. It is hugely enhanced over Lifshitz's original expectation for three reasons: the correlation-induced $\gamma$ band renormalization, the non-linear Poisson's ratio effect contributing to Young's modulus and the value of $\alpha$ in the deformation potential. In our model, the experimentally observed Van Hove strain yields $\alpha = 7.6$, and it enters $A$ as $\alpha^2$. To investigate further, we performed first-principles calculations. These emphasize that the $\gamma$ band of $Sr_2RuO_4$ is based on Ru-O-Ru processes involving two *d-p* orbital hops, yielding $\alpha = 8$ (see Supplementary Materials and Fig. S13).

Our findings also give new perspectives on the nature and consequences of the entanglement between elastic and electronic degrees of freedom in metallic solids. To what extent might they be a driver for superconductivity as a route to avoid divergent lattice softening? Does physics directly following from conduction electron density of states play a bigger than previously appreciated role in effects in heavy fermion physics such as the Kondo volume collapse (*34*) and lattice softening associated with magnetism (*35*) and metamagnetism (*36*)? While these remain open questions, our observations provide strong and concrete evidence of relevance to the 'chicken and egg' problem discussed in the introduction: conduction band physics can drive surprisingly large structural effects, and conduction electrons are not always slaves to the lattice.

**Acknowledgments:**

**Funding:**

Max Planck Society

Research in Dresden benefits from the environment provided by the DFG Cluster of Excellence ct.qmat EXC 2147, project ID 390858940 (APM)

German Research Foundation TRR 288-422213477 ELASTO-Q-MAT, Project A10 (HMLN, APM, CWH)

German Research Foundation TRR 288-422213477 ELASTO-Q-MAT, Project A11 (MG)

German Research Foundation TRR 288-422213477 ELASTO-Q-MAT, Project B01 (JS, VS)

Alexander von Humboldt Foundation Research Fellowship for Postdoctoral Researchers (HMLN)

Japan Society for the Promotion of Science Overseas Research Fellowships (KI)

KAKENHI Grants-in-Aid for Scientific Research grant no. 17H06136 (NK)

KAKENHI Grants-in-Aid for Scientific Research grant no. 18K04715 (NK)

KAKENHI Grants-in-Aid for Scientific Research grant no. 21H01033 (NK)

KAKENHI Grants-in-Aid for Scientific Research grant no. 22K19093 (NK)

Japan Society for the Promotion of Science Core-to-Core Program no. JPJSCCA20170002 (NK)

JST-Mirai Program grant no. JPMJMI18A3 (NK)

National Research Foundation of Korea grant no. 2021R1C1C1007017 (BK)

National Research Foundation of Korea grant no. 2022M3H4A1A04074153 (BK)

**Author contributions:**

Conceptualization: HMLN, JS, CWH

Formal analysis: HMLN, MG, JS, CWH

Funding acquisition: MG, JS, APM, CWH

Investigation: HMLN, KI, YSL, EG, VS, BK, IIM, MG, JS

Methodology: HMLN, EG, VS, JS, CWH

Project administration: HMLN, APM

Resources: NK, DAS

Supervision: MN, IIM, JS, APM, CWH

Visualization: HMLN

Writing – original draft: HMLN, MG, JS, BK, IIM, APM, CWH

Writing – review & editing: HMLN, KI, YSL, EG, NK, MN, BK, IIM, MG, JS, APM, CWH

**Competing interests:** CWH has 31% ownership of Razorbill Instruments, a company that markets uniaxial pressure cells. All other authors declare no competing interests.

**Data and materials availability:** All data are available in the main text or the supplementary materials.

**Supplementary Materials**

Materials and Methods

Supplementary Text

Figs. S1 to S13

Tables S1 to S2

References (*37–50*)

**Figures and Legends**

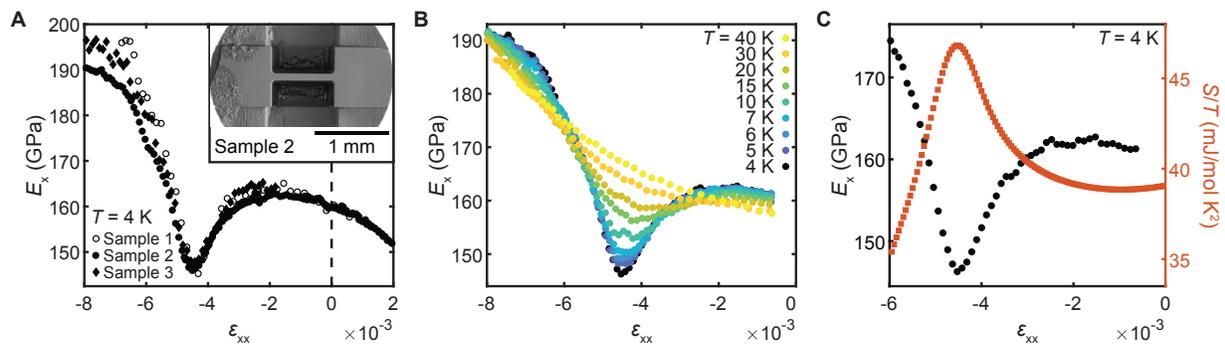

**Fig. 1. The Young's modulus $E_x$ of Sr$_2$RuO$_4$ measured across a stress-tuned Lifshitz transition.** (**A**) $E_x$ as a function of strain $\varepsilon_{xx}$ measured at 4 K on three samples. (Inset) Scanning electron micrograph of Sample 2. (**B**) $E_x$ vs. $\varepsilon_{xx}$ at a series of temperatures, measured on Sample 2. (**C**) $E_x$ at 4 K taken from the temperature series in (B) (black) together with the entropy $S/T$ extracted from elastocaloric data from a separate sample at 4 K (orange), plotted as a function of $\varepsilon_{xx}$.

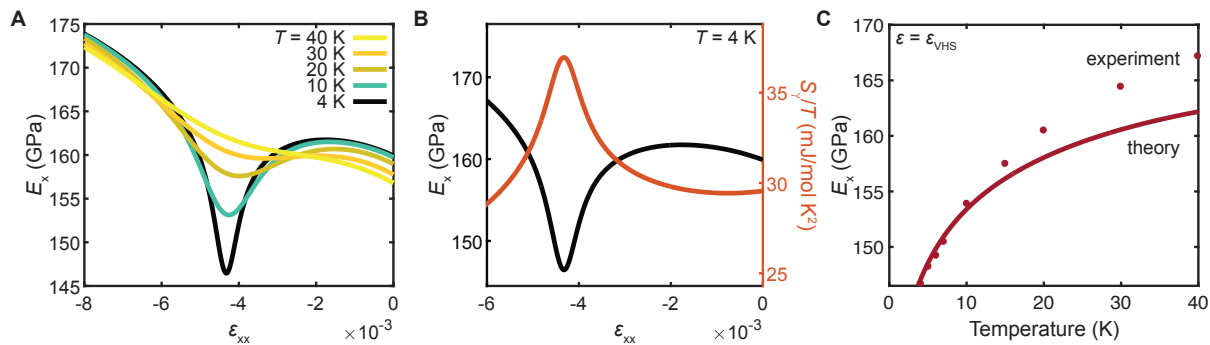

**Fig. 2. A simple model quantitatively reproduces key experimental features.** (**A**) $E_x$ vs. $\varepsilon_{xx}$ at selected temperatures. (**B**) $E_x$ and $S_\gamma/T$ vs. $\varepsilon_{xx}$ at 4 K, where $S_\gamma$ is the calculated contribution to the entropy of the $\gamma$ band. (**C**) $E_x$ vs. temperature calculated at the Van Hove strain (solid line) and corresponding experimental data (filled circles) extracted from the temperature series shown in Fig. 1B. The averaging window for the data is a strain range of $3\times10^{-4}$.

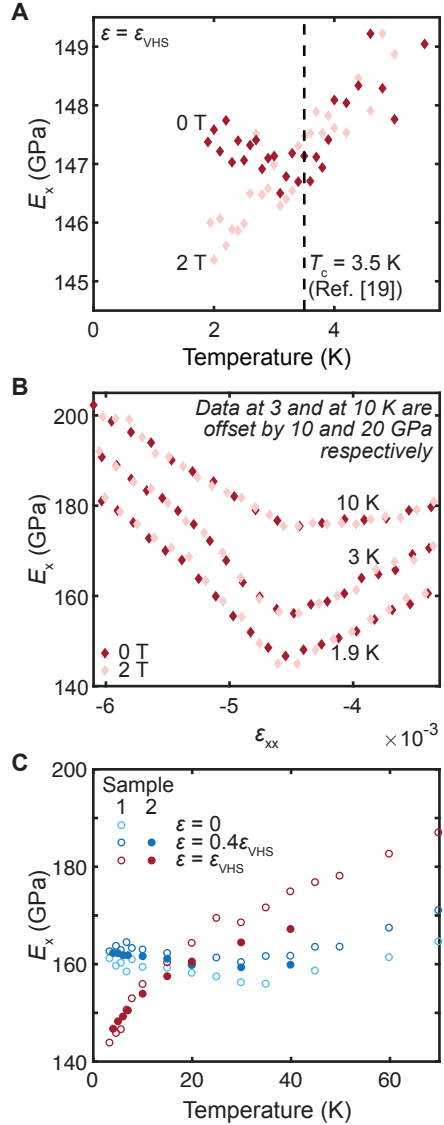

**Fig. 3. Tracking $E_x$ at key strains to lower and higher temperatures.** (**A**) $E_x$ at $\varepsilon_{VHS}$ continues to soften with decreasing temperature so long as superconductivity is suppressed with a 2 T magnetic field (pink diamonds); in the presence of superconductivity, $E_x$ hardens instead (0 T, red diamonds). The averaging window is a stress range of 0.05 GPa. (**B**) $E_x$ vs. $\varepsilon_{xx}$ at selected temperatures with and without an applied magnetic field (red, 0 T; pink, 2 T). The hardening is visible in the data at 1.9 K. The curves at 3 K and 10 K have been offset by 10 GPa and 20 GPa for clarity. Data in panels A and B are from Sample 3. (**C**) $E_x$ vs. temperature at $\varepsilon_{VHS}$ (red), $0.4\varepsilon_{VHS}$ (dark blue), and zero strain (light blue) over a wide range of temperatures. Open circles, Sample 1; filled circles, Sample 2. Data at $\varepsilon_{VHS}$ for Sample 2 are replotted from Fig. 2C. The averaging window is a strain range of $4.05\times10^{-4}$ for Sample 1 and $3\times10^{-4}$ for Sample 2.

# Supplementary Materials for

Giant lattice softening at a Lifshitz transition in $Sr_2RuO_4$

Hilary M.L. Noad, Kousuke Ishida, You-Sheng Li, Elena Gati, Veronika C. Stangier, Naoki Kikugawa, Dmitry A. Sokolov, Michael Nicklas, Bongjae Kim, Igor I. Mazin, Markus Garst, Jörg Schmalian, Andrew P. Mackenzie, Clifford W. Hicks

Correspondence to: Hilary.Noad@cpfs.mpg.de, joerg.schmalian@kit.edu, Andy.Mackenzie@cpfs.mpg.de, C.Hicks.1@bham.ac.uk

**This PDF file includes:**

    Materials and Methods
    Supplementary Text
    Figs. S1 to S13
    Tables S1 to S2

**Materials and Methods**

Sample preparation

Single-crystal Sr$_2$RuO$_4$ was grown by the optical floating zone method (*37*), cut into oriented slices using a wiresaw, and cleaved on (001) planes. The resulting bars were thinned to their final thickness, listed in Table S1, using aluminum oxide lapping film. The elastocaloric sample was carefully polished to reduce inhomogeneity and mounted directly into a strain cell without using a platform (*23*). Stress-strain samples were milled into a necked shape using a Xe plasma focused ion beam (FEI Helios G4), Figs.1A & S1A, B. We mounted the necked samples on titanium carriers using Stycast 2850/Catalyst 23LV and titanium foils to encapsulate and anchor the ends of the sample to the carrier, Fig.S1C. The carriers incorporate flexures that allow motion in the desired direction and attenuate inadvertent transverse and twisting forces, Fig. S1D. The three stress-strain samples presented in this work were taken from the same parent crystal but differed in their final dimensions, most notably in the ratio between the neck and anchoring tab widths. We summarize key dimensions in Table S1.

Extracting response of sample

We describe now how stress-strain relationships are extracted from the directly measured quantities, applied force and displacement. The complete system – sample, carrier, and cell – is approximated as a set of discrete springs joined by rigid linkages, as illustrated in Fig. S2A. The mounted sample is represented as two springs in series, one for the neck, and one for the anchoring regions, meaning the tabs of the sample itself and the epoxy and titanium foils that encapsulate the tabs. The anchors are taken to have a linear, temperature-independent force-displacement relationship defined by the spring constant $k_{anchor}$, while the neck is taken to have a general stress-strain relationship $\sigma = \sigma_{xx}(\varepsilon_{xx})$. We term this the "two-spring approximation," and its essence is to neglect regions of intermediate strain at the roots of the neck. In parallel with the sample and anchors is another spring, with spring constant $k_{flex}$, that represents the flexures of the carrier. We measure $k_{flex}$ of the empty carrier in-situ at the end of an experiment after intentionally fracturing the sample. For Samples 1 and 2, $k_{flex} \approx 0.02$ N/µm; for Sample 3, $\approx 0.05$ N/µm. The force sensor comprises a capacitive displacement sensor in parallel with a spring of spring constant $k_{force}$, which, as described in a later section, we determine separately.

In the two-spring approximation, $E_x$ is given by

$$E_x = \frac{L_{neck}}{A_{neck}} \left( \frac{1}{dF/dD} - \frac{1}{k_{anchor}} \right)^{-1} \quad \text{(Eq. S1)}.$$

$F$ is the measured force minus the force that goes into the flexures of the sample carrier, and $D$ is the measured displacement. The longitudinal strain in the sample neck, $\varepsilon_{xx}$, is given by

$$\varepsilon_{xx} = \frac{1}{L_{neck}} \left( D - \frac{F}{k_{anchor}} \right) \quad \text{(Eq. S2)},$$

where $L_{neck}$ and $A_{neck}$ are the length and cross-sectional area of the neck. $A_{neck}$ is easily measured. The choice of $L_{neck}$ is ambiguous, because there are fillets at the roots of the neck (Fig. S2B). (These fillets are necessary to limit stress concentration.) We find through simulation, discussed below, that the choice of $L_{neck}$ that yields the most accurate reconstruction of the intrinsic stress-strain relationship is the length of the straight portion of the neck, that is, excluding the fillets.

Once $L_{neck}$ is specified, we can use $dF/dD$ measured at zero force and prior knowledge of the Young's modulus $E_x$ at zero stress, 160 GPa, measured with resonant ultrasound spectroscopy (*18*), to obtain $k_{anchor}$: $dF/dD|_{F=0} = (1/k_{neck} + 1/k_{anchor})^{-1}$, where $k_{neck} = E_x(\sigma=0) A_{neck}/L_{neck}$. As force is increased, the Hooke's-law spring constant is then used to subtract off the portion of the applied displacement that goes into the tabs and epoxy, allowing accurate determination of the stress-strain relationship of the neck portion. In other words, our technique allows measurement of changes in the Young's modulus of the sample as stress is applied, starting from a fixed point at zero stress that is measured by other means.

We keep $L_{neck}$ and $A_{neck}$ constant throughout our analysis, without correction for the actual strain in the sample. These corrections would be $A(\varepsilon_{xx}) = A_{neck}(1-v_{xy}\varepsilon_{xx})(1-v_{xz}\varepsilon_{xx})$ and $L(\varepsilon_{xx}) = L_{neck}(1+\varepsilon_{xx})$. When these corrections are included, we find that at 4 K and $\varepsilon_{xx} = \varepsilon_{VHS}$, where we expect nonlinearities in $v_{xy}$ and $v_{xz}$ to be the most severe, the correction to $E_x$ is less than 1%.

Applying the two-spring analysis to Sample 1, we find $k_{anchor} = 2.7$ N/μm with $k_{neck} = (A_{neck}/L_{neck})E_x(\sigma=0) = 2.8$ N/μm at 4 K. For Sample 2, we find $k_{anchor} = 2.1$ N/μm with $k_{neck} = 2.0$ N/μm. In both cases, the anchors have a comparable stiffness to the sample neck, confirming that their deformation under applied force cannot be neglected.

Sample 3 broke during initial cooling. Data could still be collected by pushing the two parts of the sample together, but we could not obtain reliable measures of $dF/dD|_{F=0}$ because they came together somewhat gradually. Therefore, for Sample 3 we obtained $k_{anchor}$ by setting $E_x$ at the Lifshitz minimum to $E_x(\varepsilon_{VHS})$ obtained from Sample 2. This procedure yielded $k_{anchor} = 2.5$ N/μm at 4 K, similar to $k_{anchor}$ for Samples 1 and 2.

To find a good criterion for $L_{neck}$, and to test the accuracy of the two-spring model, we generated simulated data through finite element analysis in COMSOL and applied our two-spring analysis to the result. Fig. S2B shows the shape of the simulated sample. Force is applied and displacement measured over the curved end faces. For the simulations, we specified a strain-dependent Young's modulus $E_{input}(\varepsilon)$ with a sharp dip at $\varepsilon = -0.005$, similar to that seen in Sr$_2$RuO$_4$, which we plot as the black "input curve" in Fig. S2C, D. In Fig. S2C, $E_{input}(\varepsilon)$ then rises to 200 GPa with further compression, while in Fig. S2D, $E_{input}(\varepsilon)$ recovers only to its zero-stress value of 160 GPa. For simplicity, the test material is taken to be isotropic with a Poisson's ratio of 0.3.

The best match of the reconstruction to the input Young's modulus in the vicinity of the dip is obtained with $L_{neck}$ set to the length of the straight portion of the neck, i.e. excluding the fillets. We use this definition of $L_{neck}$ in our data analysis. Under further compression, the reconstructed $E_x$ in Fig. S2C exceeds $E_{input}(\varepsilon)$ by several percent. This overshoot does not appear in Fig. S2D. The overshoot in Fig. S2C is an artefact of holding $k_{anchor}$ constant at its zero-stress value: as the sample is compressed, a steadily greater portion of the anchoring tabs is compressed enough for its Young's modulus to climb to 200 GPa. In effect, the anchoring tabs stiffen, but because $k_{anchor}$ is held constant in the two-spring approximation, this stiffening appears instead as an overshoot in the reconstructed $E$.

The increase in $E$ seen in our experimental data beyond the Lifshitz transition is much larger than this overshoot from the two-spring analysis, which shows that it is a real effect. A more sophisticated analysis could in principle yield a better reconstruction of $E(\sigma)$ but would require more input parameters.

Checking assumptions about anchors

Our assumption that $k_{anchor}$ is independent of temperature is not obviously reasonable, because the anchors contain epoxy and the elastic moduli of organic compounds can have strong temperature dependence. However, the Young's modulus of the epoxy Stycast 1266 has been found to have weak temperature dependence below 77 K (*38*), and our measured $E_x(\varepsilon_{xx}=0)$ (Main Text Fig. 3C, open pale blue circles) has a temperature dependence that is not plausibly explainable as an artefact of temperature dependence of the elastic properties of the epoxy. In particular, $E_x(\varepsilon_{xx}=0)$ has a minimum at around 40 K, then increases as the temperature is raised further. If that increase were an artefact of the epoxy, it would indicate that the epoxy stiffens with increasing temperature, which is not expected.

We checked for signs of non-elastic deformation by comparing datasets collected under increasing compression (Fig. S3, grey circles) and decreasing compression (Fig. S3, black circles). In a 4 K dataset covering the largest range of compressions explored in the measurement, we find that the two directions do not lie on top of each other for $|\varepsilon_{xx}| \gtrsim 0.0065$, corresponding to $|\sigma_{xx}| \gtrsim 1$ GPa. Earlier work (*20*) found that the low-temperature elastic limit of Sr$_2$RuO$_4$ is above 2 GPa. We therefore believe that this high-stress hysteresis is not related to non-elastic deformation of the sample. Indeed, subsequent sweeps to lower maximum stresses did not exhibit the high-stress offset seen in the large-range, 4 K sweep, even at higher temperatures (for example, Fig. S3, 30 K data). We therefore conclude that the hysteresis seen at strong compression is a consequence of non-elastic deformation of the Stycast epoxy used to anchor the sample into the stress-strain cell. For consistency, we only present data collected under conditions of decreasing compression in the Main Text.

Converting capacitances to force and displacement

We convert the capacitances that we measure in the stress-strain rig to forces or displacements using the parallel-plate capacitor formula. The force on the sample is given by

$$F = \alpha_F \left( \frac{1}{C_F - C_{F,offset}} - \frac{1}{C_{F,0}(D) - C_{F,offset}} \right) \quad \text{(Eq. S3)},$$

and the total displacement is given by

$$D = \alpha_D \left( \frac{1}{C_D - C_{D,offset}} - \frac{1}{C_{D,0} - C_{D,offset}} \right) \quad \text{(Eq. S4)}.$$

$C_F$ and $C_D$ are the force and displacement capacitances that we record during a measurement. $C_{F,0}$ and $C_{D,0}$ are zero-force and zero-displacement reference values that we obtain in-situ. $C_{F,offset}$ and $C_{D,offset}$ come from the wiring and capacitor construction within the rig and are obtained from room-temperature characterizations performed separately. Previous work on a similar device (*20*) found $C_{F,offset}$ and $C_{D,offset}$ to be essentially independent of temperature. The parameters $\alpha_F$ and $\alpha_D$ are obtained from room-temperature calibrations. Conceptually, $\alpha_D = \varepsilon_0 A_{C,D}$, where $A_{C,D}$ is the displacement capacitor area, and $\alpha_F = \varepsilon_0 A_{C,F} k_{force}$, where $A_{C,F}$ is the force capacitor area and $k_{force}$ is the spring constant of the force sensor (see Fig. S2A).

We obtain the zero-force capacitance $C_{F,0}(D)$ by measuring $C_F$ vs. $C_D$ of the empty carrier after thoroughly fracturing the sample at the end of an experiment. It has a dependence on $D$ due to the flexures in the sample carrier, which continue to transmit some force to the force sensor. We fit $C_{F,0}(D)$ using a second-order polynomial and use this for the subtraction. For 10 K and above, we measure and subtract $C_{F,0}(D)$ at each temperature; for all temperatures below 10 K, we use the 4 K value. This subtraction removes the force in the flexures from the measured force. To determine the zero-displacement capacitance, $C_{D,0}$, we take the value of $C_D$ for which $C_F(D) - C_{F,0}(D) = 0$ with the sample intact. We use $C_{D,0}$ obtained from a 4 K measurement for all temperatures.

For Sample 3, which broke during initial cooling, we took $C_{D,0}$ to be the value that aligns the minimum in $E_x$ vs. strain to that of the other two samples. We determined $C_{D,0}$ for Sample 3 while using Van Hove-adjusted force calibrations for all datasets (see discussion below.)

We obtain the offset capacitances $C_{D,\text{offset}}$, $C_{F,\text{offset}}$ and parallel-plate parameters $\alpha_D$, $\alpha_F$ from room-temperature characterizations. For the displacement capacitor, we use an interferometer to independently measure the displacement of the moving block (Fig. S1D) as we apply voltage to the piezos and monitor the displacement capacitance $C_D$. We extract $C_{D,\text{offset}}$ and $\alpha_D$ by fitting the characterization data to

$$C_D = C_{D,\text{offset}} + \frac{\alpha_D}{d - d_0} \qquad (\text{Eq. S5}),$$

where $d$ is the displacement measured with the interferometer and $d_0$ is an offset, allowed to be free in the fit. For the force sensor, we take

$$C_F = C_{F,\text{offset}} + \frac{\alpha_F}{f - f_0} \qquad (\text{Eq. S6}),$$

where $f$ is the applied force and $f_0$ is an offset, allowed to be free in the fit. We obtain force sensor calibration data at room temperature by hanging calibrated weights from the stress cells.

We note that Sample 3 was measured in a commercially available cell (Razorbill UC200), and so we used the calibration parameters supplied with the cell.

We tested two different methods for obtaining the force-sensor calibration for low temperatures. In the first method, we fit the calibration data using Eq. S6 to obtain $f_0$ and $\alpha_F$, then multiply $\alpha_F$ by 1.1 as an estimate for the increase in the elastic moduli of titanium at low temperature (*20*).

In the second, we take the fitted value of $f_0$ but then ignore the fitted value of $\alpha_F$ and instead take the value of $\alpha_F$ that sets the Lifshitz stress to be -0.7 GPa. The reason to do this is that we expect the Lifshitz stress to be the same for each sample, and by imposing this constraint the shape of the minima in the Young's moduli can be compared. We summarize the calibration parameters that we used to analyze our data in Table S2.

The Young's modulus calculated with both methods of calibration is shown in Fig. S4. With the independent calibrations (Fig.S4, A and B), the Lifshitz stress is found to be -0.6 GPa in Sample 1 and -0.8 GPa in Sample 2. This variation gives a sense of the current level of uncertainty in measurement of the Lifshitz stress.

More details on data processing and extracting $E_x$ at a particular strain

In our experiment, we collect force and displacement capacitance data with a high point density at fixed temperature. We calculate force $F$ and displacement $D$ from the raw data using Eq. S3 and S4. An example of $F$ vs. $D$ is shown in Fig. S5A. Before calculating $dF/dD$ for obtaining $E_x$ (Eq. S1), we reduce the noise on our data through an averaging procedure: we divide the data into bins of fixed number of points and then average within each bin, Fig. S5B. For Samples 1 and 2, we averaged every 200 points; for Sample 3, which had a lower point density in the raw data, every 12. (Note that this is not a running average; it is essentially a down-sampling.) We then calculate $dF/dD$, $E_x$ and $\varepsilon_{xx}$ at each temperature from the binned-and-averaged $F$ and $D$ values. The points in all of the $E_x$ vs. $\varepsilon_{xx}$ curves (Figs.1A-C, 3B, S3, S4, S7) were calculated in this way.

To obtain $E_x$ at a fixed value of strain as a function of temperature (Figs. 2C, 3A, 3C), we start from the $E_x$ vs. $\varepsilon_{xx}$ data calculated as described above. We then average $E_x$ at a given temperature within a small window centered on a target strain or stress. Aside from averaging within the fixed window, no other smoothing or interpolation is applied. In Fig. S5C-E, we overlay the averaging windows on examples of the $E_x$ vs. $\varepsilon_{xx}$ data that we used to obtain the $E_x$ vs. $T$ of Figs. 2C, 3A, 3C.

For the data from Samples 1 and 2 (Figs. 2C, 3C), we defined the averaging windows in terms of strain, using $\varepsilon \pm 2.025e-4$ for Sample 1 and $\varepsilon \pm 1.5e-4$ for Sample 2, where $\varepsilon = 0$, $0.4\ \varepsilon_{VHS}$, or $\varepsilon_{VHS}$. For the data from Sample 3 (Fig. 3A), we defined the averaging window in terms of stress, using $\sigma_{VHS} \pm 0.025$ GPa.

The points for Sample 1 shown in Fig. 3C came from a temperature series of which the curve in Fig. 1A was a part; the points for Sample 2 in Figs. 2C and 3C were obtained from the temperature series shown in Fig. 1B; and the points for Sample 3 in Fig. 3A came from the temperature series that is partially shown in Fig. 3B.

**Supplementary Text**

Further theory details

*Tight-binding model*

To develop a quantitative understanding of the measured elastic properties of Sr$_2$RuO$_4$, we use a model for the $\gamma$ band that was recently employed to describe the strain and temperature dependence of the elastocaloric effect in Sr$_2$RuO$_4$ (*23*). Using the ARPES data of Ref. (*25*), we model the tight-binding dispersion as

$$E_{\mathbf{k}} = -2t_x \cos(k_x a_x) - 2t_y \cos(k_y a_y) - 4t'\left(\cos(k_x a_x)\cos(k_y a_y)\right) - \mu \quad \text{(Eq. S7)}.$$

At zero strain, $a_x = a_y = a_0$, $t_x = t_y = t_0$, and $t' = t'_0$, with parameters $t_0 = 0.119$ eV, $t'_0/t_0 = 0.392$, and $\mu/t_0 = 1.48$. The values for $t_0$ and the chemical potential $\mu$ are directly taken from Ref. (*25*), while the value for the next-nearest neighbor hopping was obtained by fitting the dispersion near the Y-point of the unstrained system and is only very slightly smaller than what is given in Ref. (*25*). Under strain, $a_x = a_0(1 + \varepsilon_{xx})$ and $a_y = a_0(1 + \varepsilon_{yy})$ describe the strain dependence of the lattice constants. For the strain dependence of the hopping elements, we use the following description (*39*):

$$t_x = t_0\left(1 - \alpha \varepsilon_{xx} + \beta \varepsilon_{zz}\right) \quad \text{(Eq. S8a)},$$

$$t_y = t_0\left(1 - \alpha \varepsilon_{yy} + \beta \varepsilon_{zz}\right) \quad \text{(Eq. S8b)},$$

$$t' = t_0'\left(1 - \frac{\alpha'}{2}\left(\varepsilon_{xx} + \varepsilon_{yy}\right) + \beta' \varepsilon_{zz}\right) \quad \text{(Eq. S8c)}.$$

We chose $\alpha = \alpha' = 7.604$, which moves the Van Hove singularity at $\mathbf{k}_{VHS} = \left(\frac{\pi}{a_x}, 0\right)$ to the Fermi energy for $\varepsilon_{VHS} = 0.44\%$ (value from (*18*)). These values for $\alpha$ and $\alpha'$ yield a strain dependence of $E(\mathbf{k}_{VHS})$, the band energy at the point in $k$ space where the Lifshitz transition occurs, that is in good agreement with the first-principles calculations of Ref. (*18*). From Ref. (*40*) follows $\beta = 0.385$ and $\beta' = 1.155$.

It now follows that the free-energy density of the problem is

$$F = -2k_B T \int \frac{d^3k}{(2\pi)^3} \log\left(1 + e^{-\beta E_\mathbf{k}}\right) + \tfrac{1}{2}\hat{\varepsilon}^T \hat{C}_0 \hat{\varepsilon} - \hat{\sigma}_0^T \hat{\varepsilon} \quad \text{(Eq. S9)},$$

where the first term is the contribution due to the electronic states in the $\gamma$ band, while $\hat{\varepsilon} = \left(\varepsilon_{xx}, \varepsilon_{yy}, \varepsilon_{zz}\right)$ are the three uniaxial strain values and

$$\hat{C}_0 = \begin{pmatrix} C_{0,11} & C_{0,12} & C_{0,13} \\ C_{0,12} & C_{0,22} & C_{0,23} \\ C_{0,13} & C_{0,23} & C_{0,33} \end{pmatrix} \quad \text{(Eq. S10)}$$

is the section of the elastic constant tensor relevant for our analysis, i.e. without shear strain. $\hat{C}_0$ includes the response due to the lattice, the core electrons, and the $\alpha$ and $\beta$ sheets of the Fermi surface, i.e. of all degrees of freedom not included in the $\gamma$ band. We assume that $\hat{C}_0$ depends only weakly on strain and temperature and simply ignore those dependencies. $\hat{\sigma}_0^T = \left(\sigma_{0,x}, \sigma_{0,y}, \sigma_{0,z}\right)$ is the internal stress field that compensates the electron pressure and ensure that the crystal is in equilibrium at zero external stress.

It then follows for the elastic constant:

$$C_{ij} = C_{0,ij} - \frac{2}{k_B T} \int \frac{d^3k}{(2\pi)^3} f_{\mathbf{k}}(1-f_{\mathbf{k}}) \frac{\partial E_{\mathbf{k}}}{\partial \varepsilon_{ii}} \frac{\partial E_{\mathbf{k}}}{\partial \varepsilon_{jj}}$$ (Eq. S11),

with the Fermi function $f_{\mathbf{k}}$. The above expression is based on the fact that $\frac{\partial^2 E_{\mathbf{k}}}{\partial \varepsilon_{ii}^2} = 0$ for our model. Notice that the correction due to the $\gamma$ band is negative and will always soften the elastic modulus.

The elastic tensor of Sr$_2$RuO$_4$ at zero strain was determined in Ref. (*41*) (see also Supplementary Material of Ref. (*23*)). We determine the $C_{0,ij}$ such that $C_{ij}(\sigma_{xx}=0, T=4K)$ agrees with the measured elastic constants of Refs. (*23*). This yields

$$\hat{C}_0 = \begin{pmatrix} 268.465 & 123.225 & 82.55 \\ 123.225 & 268.465 & 82.55 \\ 82.55 & 82.55 & 258.273 \end{pmatrix}$$ (Eq. S12),

where all values are given in GPa. Notice that $\hat{C}_0$ still has the symmetry of the tetragonal state at zero stress. Stress-induced orthorhombic effects with $C_{0,11} \neq C_{0,22}$ and $C_{0,13} \neq C_{0,23}$ are, in our model, only due to the $\gamma$ band.

Finally, since we know that the stable crystal at zero external stress has no strain in equilibrium, we determine the internal stresses as

$$\sigma_{0,ii} = 2 \int \frac{d^3k}{(2\pi)^3} f_{\mathbf{k}}^0 \frac{\partial E_{\mathbf{k}}}{\partial \varepsilon_{ii}}$$ (Eq. S13),

where the dispersion in $f_{\mathbf{k}}^0$ is taken at zero strain.

*Non-linear elasticity*

Very generally, we expect to be in the non-linear regime at large external stresses. Moreover, if there is a qualitative change in behavior as a function of stress, such as when crossing a Van Hove singularity, non-linear effects become important. In the following section we show how to account for non-linear elasticity in the calculation of $E_x$ and demonstrate the impact of these effects in Sr$_2$RuO$_4$.

We start by considering the free energy as a function of strain, $\hat{F}(\hat{\varepsilon})$, where $\hat{\varepsilon} = (\varepsilon_{xx}, \varepsilon_{yy}, \varepsilon_{zz})$. As before we only consider uniaxial strain $\varepsilon_{xx}$, $\varepsilon_{yy}$, and $\varepsilon_{zz}$ and ignore shear strain. Without external stress the equilibrium state corresponds to $\hat{\varepsilon} = 0$. We include stress via

$$F \rightarrow G = F - \hat{\sigma}^T \hat{\varepsilon}$$ (Eq. S14),

where $\hat{\sigma} = (\sigma_{xx}, 0, 0)$. In other words, as in the experiment, there is no stress applied along the $y$ or $z$ directions.

Let us now consider a system where the function $F(\hat{\varepsilon})$ is not necessarily harmonic. In the actual calculation we will of course use the free energy of Eq. S9. If we start from

$$G = F(\varepsilon_{xx}, \varepsilon_{yy}, \varepsilon_{zz}) - \sigma_{xx}\varepsilon_{xx}$$ (Eq. S15),

the equations of state follow from minimization with respect to the three strain values:

$$\frac{\partial F(\varepsilon_{xx}, \varepsilon_{yy}, \varepsilon_{zz})}{\partial \varepsilon_{xx}} = \sigma_{xx}$$ (Eq. S16a),

$$\frac{\partial F(\varepsilon_{xx}, \varepsilon_{yy}, \varepsilon_{zz})}{\varepsilon_{yy}} = 0$$ (Eq. S16b),

$$\frac{\partial F(\varepsilon_{xx}, \varepsilon_{yy}, \varepsilon_{zz})}{\partial \varepsilon_{zz}} = 0 \quad \text{(Eq. S16c)}.$$

We can now solve the last two non-linear equations and determine $\varepsilon_{yy}$ and $\varepsilon_{zz}$ as a function of $\varepsilon_{xx}$:

$$\varepsilon_{yy}(\varepsilon_{xx}) = -\nu_{xy}(\varepsilon_{xx})\varepsilon_{xx} \quad \text{(Eq. S17a)},$$

$$\varepsilon_{zz}(\varepsilon_{xx}) = -\nu_{xz}(\varepsilon_{xx})\varepsilon_{xx} \quad \text{(Eq. S17b)},$$

where $\nu_{xy}(\varepsilon_{xx})$ and $\nu_{xz}(\varepsilon_{xx})$ are Poisson ratios. In linear elasticity, the Poisson ratios do not depend on strain; once the free energy has anharmonicities – that is, in the non-linear regime – the Poisson ratios may depend on strain.

Inserting $\varepsilon_{yy}(\varepsilon_{xx})$ and $\varepsilon_{zz}(\varepsilon_{xx})$ into the equations of state, Eq. S16, it follows that

$$\sigma_{xx} = \frac{\partial F(\varepsilon_{xx}, \varepsilon_{yy}(\varepsilon_{xx}), \varepsilon_{zz}(\varepsilon_{xx}))}{\partial \varepsilon_{xx}} \quad \text{(Eq. S18)}.$$

This yields for the Young's modulus

$$E_x = \frac{d\sigma_{xx}}{d\varepsilon_{xx}} = \frac{\partial^2 F}{\partial \varepsilon_{xx}^2} + \frac{\partial^2 F}{\partial \varepsilon_{yy} \partial \varepsilon_{xx}} \frac{\partial \varepsilon_{yy}}{\partial \varepsilon_{xx}} + \frac{\partial^2 F}{\partial \varepsilon_{zz} \partial \varepsilon_{xx}} \frac{\partial \varepsilon_{zz}}{\partial \varepsilon_{xx}} \quad \text{(Eq. S19)}.$$

Identifying the elastic constants as $C_{ab} = \partial^2 F / \partial \varepsilon_a \partial \varepsilon_b$ and the differential Poisson ratios as $\nu_{xy}^{(d)} = -\partial \varepsilon_{yy}/\partial \varepsilon_{xx}$ and $\nu_{xz}^{(d)} = -\partial \varepsilon_{zz}/\partial \varepsilon_{xx}$, we re-write the Young's modulus as

$$E_x = C_{11} - \nu_{xy}^{(d)} C_{12} - \nu_{xz}^{(d)} C_{13} \quad \text{(Eq. S20)}.$$

Within our model of $Sr_2RuO_4$, non-linear effects are most important at low temperatures, near the Van Hove point. In Fig. S6A, B we compare the strain dependence of the differential Poisson ratio $\nu_{xy}^{(d)}$ to its standard counterpart. At high temperatures, $T = 40$ K (Fig. S6B), only small deviations between $\nu_{xy}$ and $\nu_{xy}^{(d)}$ occur, and only when the strain becomes large. In contrast, at low temperatures, $T = 4$ K (Fig. S6A), $\nu_{xy}^{(d)}$ is strongly enhanced at the Van Hove point. As shown in Fig. S6C, D, $\nu_{xz}^{(d)}$ and $\nu_{xz}$ display similar trends to those of $\nu_{xy}^{(d)}$ and $\nu_{xy}$, except with the opposite sign – where $xy$ increases, $xz$ decreases, and vice versa. The importance of using differential Poisson ratios in obtaining $E_x$ is illustrated in Fig. S6E, F. At low temperatures, $T = 4$ K (Fig. S6E), replacing $\nu_{xy}^{(d)}$ and $\nu_{xz}^{(d)}$ with $\nu_{xy}$ and $\nu_{xz}$ reduces the dip in $E_x(\varepsilon_{xx})$ at the Van Hove point by more than a factor of two.

The central result of our calculations is shown in Fig. S7, where we plot $E_x$ calculated as a function of compressive and tensile strain at 4 K together with experimental data from Sample 2 obtained at the same temperature. In Fig. S8, we plot our calculated $E_x$ as a function of temperature at three strains, in analogy to the experimental data plotted in Fig. 3C of the Main Text.

*Scaling theory and quantum Harris criterion*

In this section we perform an analysis of the elastic constant near a strain-tuned quantum critical point with hyperscaling. We will see that under certain circumstances, when a so-called quantum Harris criterion is fulfilled, a mechanical catastrophe takes place that renders the critical point unstable (*27, 28, 42*). As will be discussed in the next section, the behavior near a two-dimensional Lifshitz point is the marginal analogue to this mechanical instability.

The starting point of our analysis is the scaling expression for the entropy

$$S(T, \varepsilon) = T^{d/z} \phi\left(\frac{\varepsilon - \varepsilon_c}{T^{1/\nu z}}\right) \quad \text{(Eq. S21)},$$

also given in Eq. 3 of the Main Text. $\nu$ is the correlation length exponent, $z$ is the dynamic scaling exponent, and $d$ is the number of dimensions. The scaling function is finite at vanishing argument. In addition, we use that the entropy is maximal at the QCP, i.e. that $\phi''(0) < 0$, with $\phi''$ the second derivative of $\phi$.

We integrate $S(T,\varepsilon) = -\partial F(T,\varepsilon)/\partial T$ between some reference temperature $T_0$ and the actual temperature and obtain

$$F(T,\varepsilon) = F(T_0,\varepsilon) - \int_{T_0}^{T} T'^{d/z} \phi\left(\frac{\varepsilon - \varepsilon_c}{T'^{1/\nu z}}\right) dT' \qquad \text{(Eq. S22)}.$$

This leads to the elastic constant $C(T,\varepsilon) = \partial^2 F(T,\varepsilon)/\partial \varepsilon^2$ at the quantum critical point

$$C(T,\varepsilon_c) = C(T_0,\varepsilon_c) - \phi''(0) \int_{T_0}^{T} T'^{\frac{\nu d - 2}{\nu z}} dT' \qquad \text{(Eq. S23)}.$$

To proceed we need to distinguish systems where the combination $\nu(d+z) - 2$ is positive or negative.

If $\nu(d+z) > 2$, the integral is infrared convergent. We can choose $T_0 = 0$ and it follows

$$C(T,\varepsilon_c) = C_0 - \frac{\phi''(0)\nu z}{\nu(d+z) - 2} T^{\frac{\nu(d+z)-2}{\nu z}} \qquad \text{(Eq. S24)}.$$

Here, $C_0$ is the ground-state contribution to the elastic constant. Since $\phi''(0) < 0$, a universal positive temperature correction occurs at the QCP which vanishes as $T \to 0$. The elastic constant softens as one decreases the temperature, but the system remains mechanically stable.

The situation changes qualitatively when the quantum Harris criterion $\nu(d+z) < 2$ is fulfilled. Now, the integral is divergent for $T_0 \to 0$. This alone demonstrates that the above scaling form for the entropy cannot be correct down to lowest temperatures. The system either crosses over to a new critical regime where strain becomes a genuine dynamical quantum-critical mode or undergoes a finite-temperature phase transition to a new state of matter. It is however possible to use $T_0 \cong T_F$, where $T_F$ plays the role of the largest energy scale where the scaling theory applies. Integration then yields the same result as Eq. S24; however, it now holds that the temperature independent constant is

$$C_0 = C(T_F,\varepsilon_c) + \frac{\phi''(0)\nu z}{\nu(d+z) - 2} T_F^{\frac{\nu(d+z)-2}{\nu z}} \qquad \text{(Eq. S25)}.$$

We still expect that $\phi''(0) < 0$ as the entropy should still be maximal at $\varepsilon = \varepsilon_c$. For $\nu(d+z) < 2$ the temperature correction in Eq. S24 is negative and divergent. Hence, the elastic constant vanishes at a finite temperature, signaling the expected mechanical instability. No strain-tuned quantum critical point is possible when the quantum Harris criterion $\nu(d+z) < 2$ is fulfilled; a mechanical catastrophe becomes inevitable. The softening of the Young's modulus in our system is the marginal precursor of this behavior.

*Scaling for marginal case*

The Lifshitz point of Sr$_2$RuO$_4$ is part of a more general class of systems where the quantum Harris criterion (*27, 28, 42*),

$$\nu(d+z) < 2 \qquad \text{(Eq. S26)},$$

plays a crucial role in affecting quantum criticality. Another example is of one-dimensional spin chains, $d=1$, at their field-induced quantum critical point (*42*). For 1d Ising criticality $\nu=z=1$, the criterion is only marginally satisfied, $\nu(d+z)=2$, but an elastic coupling can still lead to a non-trivial renormalization group flow, see for example Ref. (*43*). However, for the isotropic Heisenberg chain at its field-induced quantum critical point, with $\nu=\frac{1}{2}$ and $z=2$, the criterion of Eq. S26 is fulfilled and an instability due to an elastic coupling is expected.

The behavior near the Van Hove point in Sr$_2$RuO$_4$ is governed by a logarithmic density of states, at least in the regime where interlayer coupling effects can be ignored. It leads to a logarithmic $T$-dependence of the heat capacity coefficient $\gamma=\frac{C}{T}=\gamma_0 \log\frac{T_F}{T}$ for $T \ll T_F$. Such behavior occurs in a much more general context and is one of the key aspects of the marginal Fermi liquid (*44*). In this section we discuss the rather generic behavior as it occurs at any quantum-critical point that displays the above logarithmic temperature dependence, regardless of its microscopic origin. It is the marginal case where Eq. S26 is fulfilled with an equal sign.

We consider an electronic system with a Sommerfeld coefficient that diverges logarithmically near an electronic instability and start our analysis from the entropy:

$$S(T,\varepsilon) = \frac{\gamma_0}{2} T \log \frac{T_F^2}{T^2 + T^*(\varepsilon)^2} \quad \text{(Eq. S27),}$$

where $\gamma_0$ is a strain- and temperature-independent prefactor, $T_F$ an upper cutoff in units of temperature, and $T^* = A(\varepsilon - \varepsilon_c)$ a tuning parameter that vanishes at a critical strain value $\varepsilon_c$. In Fig. S9 we show entropy as a function of $\varepsilon$ for different $T$ (Fig. S9A) and as a function of temperature for different $T^*(\varepsilon)$ (Fig. S9B).

By integrating the entropy over $T$ we can, alternatively, obtain the free energy

$$F(T,\varepsilon) = E_0(\varepsilon) - \int_0^T S(T',\varepsilon) dT' \quad \text{(Eq. S28),}$$

where $E_0(\varepsilon)$ is some temperature-independent background contribution due to degrees of freedom that are not responsible for the singular entropy of Eq. S27, such as core electrons. Performing the integral and taking the second derivative with respect to $\varepsilon$ yields the full elastic constant

$$C_{el} = C_{el}^{(0)}(T_0) - \gamma_0 A^2 \int_{T_0}^T dT' T' \frac{T^*(\varepsilon)^2 - T'^2}{\left(T^*(\varepsilon)^2 + T'^2\right)^2} \quad \text{(Eq. S29)}$$

where $C_{el}^{(0)}(T_0) = \partial^2 F_0(\varepsilon,T_0)/\partial \varepsilon^2$ is the corresponding background contribution to the elastic constant at temperature $T_0$. For $T_0=0$ the integral is infrared divergent at the critical point $\varepsilon=\varepsilon_c$. This demonstrates a mechanical instability of any strain-tuned marginal critical point. Hence, the logarithmic behavior of the entropy of Eq. S27 behaves similarly to systems with $\nu(d+z)<2$, yet with a logarithmic correction to the ground state elastic constant.

To obtain an impression of the finite-$T$ behavior that follows from Eq. S27 we use $T_0=T_F$, which yields for not too low temperatures a behavior in qualitative agreement with what follows from our microscopic model of the $\gamma$ band. In Fig. S10 we show the behavior of the elastic constant normalized by its background value $C_{el}^{(0)}(T_F)$ as a function of strain for different temperatures (Fig. S10A) and as a function of $T$ for different strains (Fig. S10B). We use $\gamma_0=2$, $\frac{C_{el}^{(0)}(T_F)}{\gamma_0 A^2}=30$, and measure temperatures in units of $T_F$. The choice of $\frac{C_{el}^{(0)}(T_F)}{\gamma_0 A^2}$ is really only a convenient constant shift of the elastic constant. The appropriate combination of elastic constants leads to the Young's modulus $E_x$ and the conceptually equivalent Eq. 2 of the Main Text.

*Lattice softening at the Lifshitz transition in three-dimensional materials*

The softening at a Lifshitz transition in a three-dimensional metal is estimated. We assume a tight-binding dispersion with hopping parameter $t$ of a simple cubic lattice with lattice constant $a$,

$$E(\mathbf{k}) = -2t\left(\cos(k_x a) + \cos(k_y a) + \cos(k_z a)\right) \quad \text{(Eq. S30).}$$

The density of states is given by

$$\nu(\varepsilon) = \int_{-\pi/a}^{\pi/a} \frac{dk_x dk_y dk_z}{(2\pi)^3} \delta(\varepsilon - E(\mathbf{k})) \quad \text{(Eq. S31).}$$

The free energy density at constant chemical potential $\mu$ and at finite temperature is given by

$$f(\mu, t, T) = -k_B \int_{-\infty}^{\infty} d\varepsilon \nu(\varepsilon) \log\left(1 + \exp\left(-\frac{\varepsilon - \mu}{k_B T}\right)\right) = \frac{t}{a^3} \Phi\left(\frac{\mu}{t}, \frac{k_B T}{t}\right) \quad \text{(Eq. S32),}$$

where $\Phi$ is a dimensionless scaling function that depends on the chemical potential $\mu$, the temperature $T$ with the Boltzmann constant $k_B$, and the hopping amplitude $t$. We assume that the hopping depends linearly on strain $t(\varepsilon) = t_0(1 + \alpha\varepsilon)$, where the dimensionless parameter $\alpha$ quantifies the strain dependence.

In order to estimate the softening of the elastic constant at the Lifshitz transition, we first focus on the limit of zero temperature, $T = 0$. The contribution of the tight-binding electrons to the elastic constant at constant chemical potential then reads

$$\delta C = \left.\frac{\partial^2 f_0}{\partial \varepsilon^2}\right|_{\varepsilon=0} = \frac{\alpha^2}{a^3} \frac{\mu^2}{t} \Phi''\left(\frac{\mu}{t}, 0\right) \quad \text{(Eq. S33).}$$

The relative correction to the elastic constant due to the conduction electrons can then be put into the form

$$\frac{\delta C}{C} = \frac{\alpha^2 t}{a^3 C} \Psi(\mu/t) \quad \text{(Eq. S34).}$$

The dimensionless function $\Psi(x) = x^2 \Phi''(x, 0)$ is shown in Fig. S11A. It is negative, amounting to a softening of the crystal lattice. There are two Lifshitz transitions at $x = \pm 6$ where the tight-binding band is either at the edge of getting completely empty or filled. Two saddle-point Lifshitz transitions, which are of interest to us, are located at $x = \pm 2$. Here, the cusp induces a drop of the dimensionless function on the order of 0.04. In order to estimate the relative change $\frac{\delta C}{C}$ close to a three-dimensional saddle-point Lifshitz transition we take $t \sim 0.2$ eV, $a \sim 4$ Å, $C \sim 200$ GPa and $\alpha \sim 1$ to obtain $\frac{\alpha^2 t}{a^3 C} \sim 0.0025$. Taking into account the variation of the dimensionless function $\Psi(x)$ we obtain the estimate $\frac{\delta C}{C} \sim 10^{-4}$ cited in the Main Text. For a strain dependence with $\alpha \sim 10$ the estimate is instead on the percent level, $\frac{\delta C}{C} \sim 10^{-2}$. This analysis was performed at constant chemical potential. The analysis of this single-site tight-binding band at constant density even yields $\delta C = 0$, implying that our estimate serves as an upper bound.

In Fig. S12 we also show the resulting entropy per unit cell and the change in the elastic constant. Notice, compared to a background constant of 200 GPa these changes are indeed of the mentioned order of magnitude. The relative change in the entropy are larger than the changes in the elastic constants. However, analysing the derivative in the entropy we find that $\partial^2 S/\partial \varepsilon^2$ does indeed equal $-\partial C/\partial T$ as it has to.

In Fig. S11B we show, for comparison, the corresponding scaling function $\Psi(x)$ for a two-dimensional problem. Since the effect of the chemical potential-tuned transition vanishes in $d = 2$ if the density of states is particle hole symmetric, we allow for a small next nearest neighbor hopping, see Eq. S7. We see that the anomaly in $\Psi(x)$ is significantly larger and of the order of 0.4, yielding $\frac{\delta C}{C} \sim 10^{-3}$ for a deformation potential parameter $\alpha = 1$. For $\alpha = 10$ we do find $\frac{\delta C}{C} \sim 10^{-1}$. This reveals that the observed large softening is the consequence of two effects: i) the two-dimensionality with its logarithmic density of states, which yields an enhancement of factor 10 compared to $d = 3$ and ii) the large deformation potential of $\alpha \approx 8$ which yields, with $\delta C \propto \alpha^2$, a factor of order $10^2 t$, regardless of dimension. As discussed, this value for the deformation potential is what one expects for overlapping 4d and 2p orbitals, while $\alpha = 1$ for s-orbitals.

## Density Functional Theory (DFT) results

### Calculation details

Electronic structure calculations were performed employing the projector-augmented-wave method implemented in the Vienna ab initio simulation package (VASP), where we have used the generalized gradient approximation (GGA) by Perdew-Burke-Ernzerhof functional (*45-47*). An energy cut-off of 600 eV is used for the plane-wave expansion with 40 × 40 × 20 Monkhorst-Pack $k$-mesh sampling. To obtain the series of the strained structures, we fixed the $a$ lattice parameter and fully optimized all the other lattice parameters and internal degrees of freedom. Due to the innate over-binding tendency of the GGA functional, the critical stress of the Lifshitz transition is overestimated: our calculations report $\sigma = 1.8$ GPa ($a$ of 3.828 Å). Despite the differences, however, as we have shown in our previous report (*48*), the evolution of the electronic structure and magnetic properties is well-captured before and beyond the Lifshitz transition.

### Stress

A softening of Young's modulus is readily visible in the DFT calculations. In Fig. S13, we plot $E_x$ as a function of strain, calculated from the finite difference ratio $E_x = \Delta\sigma/\Delta\varepsilon$, where $\Delta$ denotes the difference between the calculated stress value at a given lattice parameter $a$, and the value at the calculated unstrained lattice parameter $a_0$. Note that there is an uncertainty in implementing this protocol: if instead of the calculated $a_0$ we would have taken the experimental one, the observed softening would deviate by a few tenths of a percent. Note also that this protocol is different from the procedure used for the experimental data, in which the derivative of $F$ with respect to $D$ is taken as a point-by-point gradient.

As the strain is applied, there is an expected increase of the Young's modulus starting from the unstrained case. One can see a dip (softening) around the Lifshitz transition. The calculated softening is about twice as small as the experimental one. Given the subtlety of the calculated property and the general inaccuracy of the GGA-DFT, this is an acceptable error. Also, due to GGA over-binding, the calculated critical strain is, as discussed, 1.3%, compared to the experimental value of 0.4%. One can translate this into a smaller softening. A final word of caution is that, while spin-orbit coupling and correlations beyond DFT are not included in the current calculation, they only represent small corrections to the DFT total energy. However, they do generate a shift of the position of the Van Hove singularity with respect to the Fermi energy (*18*). Still, it is important to note that our parameter-free DFT calculations qualitatively and even semi-quantitatively describe the observed modulus softening.

*Connection with the model calculations*

*Force theorem.* Self-consistent energies calculated in DFT include, besides the one-electron energy, $E_1$ (that is, the sum of energies of all occupied electron states), contributions from the Hartree energy, electron-ion interaction, and the DFT exchange-correlation energies. Why than can we successfully simulate these energy differences in our model that only includes $E_1$? The answer comes from the so-called Andersen's force theorem (*49*), which states that if a self-consistent DFT charge density is perturbed by a weak external potential, the difference between the two *self-consistent* total energies is equal, to the second order in perturbation, to the difference of the *one-electron energies*, calculated with the *same non-self-consistent density*, but explicitly applying the perturbed external potential. In other words,

$$E_{SCF}[\rho_1, V_1] - E_{SCF}[\rho_0, V_0] \approx E_1[\rho_0, V_1] - E_1[\rho_0, V_0] \quad \text{(Eq. S35)}.$$

The right-hand side of Eq. S35 is exactly what we do in our analytical model.

*Canonical scaling*. Finally, using the same force theorem, we can estimate the dependence of our key parameter, the hopping amplitude, which determines the scale of the DFT band energies, on strain, i.e., on the bond length. Indeed, it was shown by Andersen and collaborators (*50*) that in the KKR methods and its derivatives (LMTO, ASW), the overlap integrals between the atomic orbitals with the angular momentum $l$ and $l'$ scales as $1/r^{l+l'+1}$, where $r$ is the bond length. Given that the $d$-$d$ hopping in the compound is indirect, via oxygen $p$ orbitals, and integrating out the latter (which is, of course, an approximation), we obtain $t_{dd} \sim t_{pd}^2/\Delta$ with charge-transfer energy $\Delta$. Since $l=2$ and $l'=1$, we obtain $t_{pd} \sim r^{-4}$, such that $t_d \sim r^{-8}$ for the effective $d$-$d$ hopping. Hence, the coefficient $\alpha = \frac{-\partial \log(t_{dd})}{\partial r} = 8$, very near the value $\alpha \simeq 7.6$ of the low-energy model of the $\gamma$ band that yields the correct value for the strain value of the Lifshitz transition. This result demonstrates that the sensitivity of the electronic structure in transition-metal oxides to strain is significantly larger than that of bands dominated by $s$-orbitals, for which $\alpha=1$. Since the corrections to the elastic constants go as $\alpha^2$, this difference makes an important contribution to the magnitude of the softening in $Sr_2RuO_4$.

**Fig. S1.**

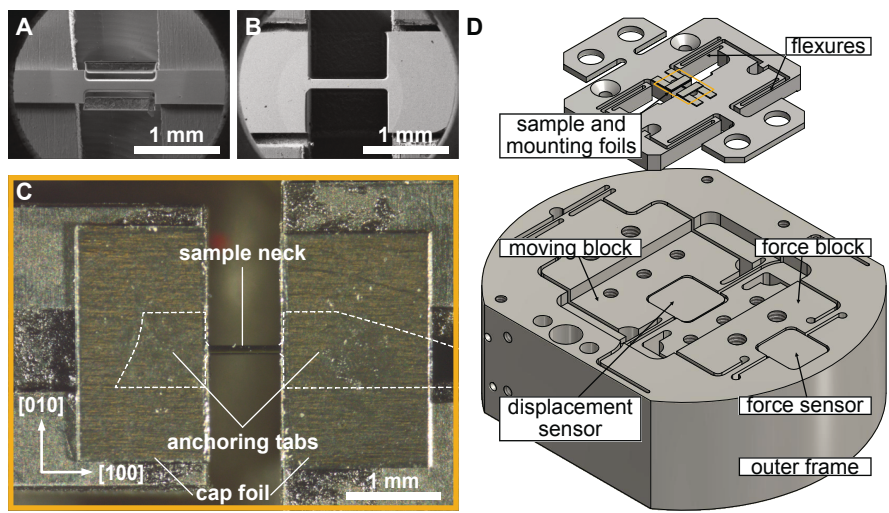

**Preparing and mounting samples for stress-strain measurements in a piezo-driven cell.** (**A, B**) Additional FIB-sculpted samples. A: Sample 1; B, Sample 3. (**C**) Optical image of Sample 2 mounted in epoxy and titanium foils on a titanium carrier. (**D**) Schematic of the piezo-driven stress-strain cell with the sample carrier and sample. The yellow outline in D indicates the approximate location of the area that is framed in yellow in C.
28

**Fig. S2.**

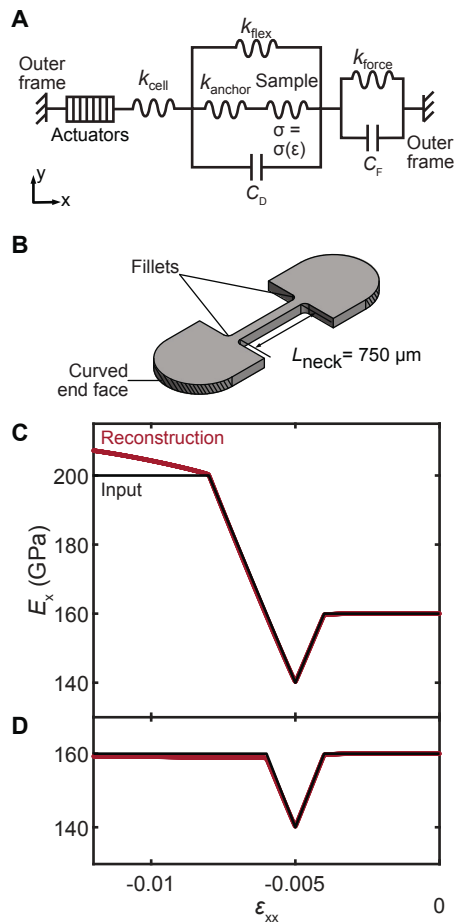

**Modeling the sample-carrier-cell system in order to reconstruct the Young's modulus and strain of the sample neck from the measured total force and displacement.** (**A**) Discretized model of the stress-strain cell and the sample. (**B**) Drawing of the model sample used in finite-element simulations. (**C, D**) Illustration of artefacts introduced by the two-spring analysis. The black lines are the input Young's moduli for finite-element analysis of the sample deformation, and the red lines are the Young's moduli reconstructed from the sample deformation using the two-spring analysis.

**Fig. S3.**

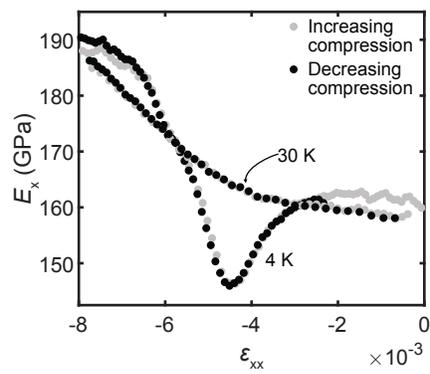

**Checking for signs of non-elastic deformation at high compression.** Grey points, increasing compression; black points, decreasing compression. Data shown here are from Sample 2. The 4 K, decreasing-compression data are also plotted in Fig. 1A of the Main Text, and the 30 K, decreasing compression data in Fig. 1B.

**Fig. S4.**

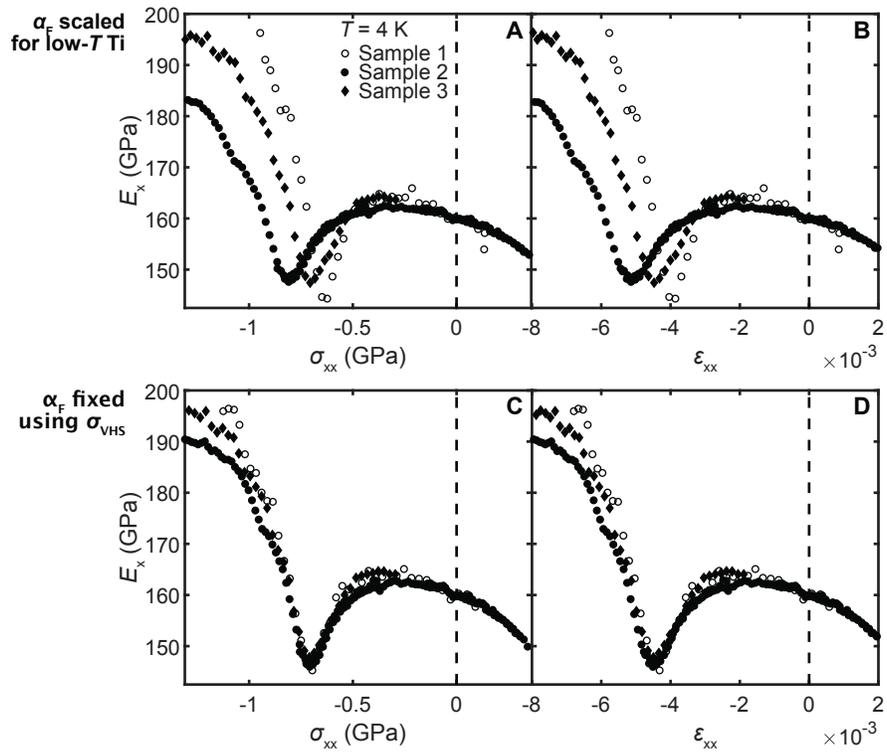

**Examining the effects of a leading source of systematic uncertainty, the force sensor calibration constant $\alpha_F$.** 4 K datasets of Fig.1A analyzed using independently-fitted values of $\alpha_F$ (**A, B**) and using values of $\alpha_F$ that set the Lifshitz stress to -0.7 GPa (**C, D**). Data are plotted against stress in panels A and C, and against strain in panels B and D.

**Fig. S5.**

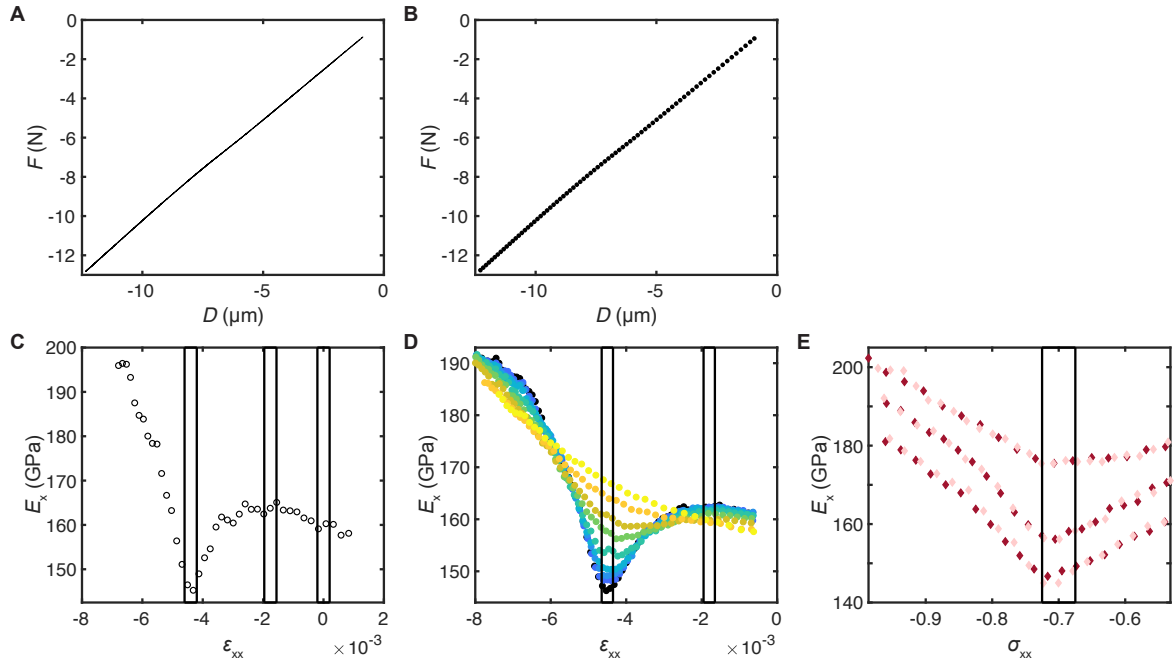

**Obtaining $E_x$ as a function of temperature at selected strains.** Force and displacement data, calculated from measured capacitances, (**A**) before and (**B**) after binning and averaging. The force-displacement data shown here are the 4 K data from Sample 2 that underly the plot in Fig. 1B. After calculating $E_x$ and $\varepsilon_{xx}$ from force and displacement, we extract $E_x$ at a particular strain (or stress) by taking the average of $E_x$ in a small window. Averaging windows overlaid on data from the Main Text for (**C**) Sample 1 (from Fig. 1A), (**D**) Sample 2 (Fig. 1B), and (**E**) Sample 3 (Fig. 3B, here plotted as a function of $\sigma_{xx}$ rather than $\varepsilon_{xx}$).

**Fig. S6.**

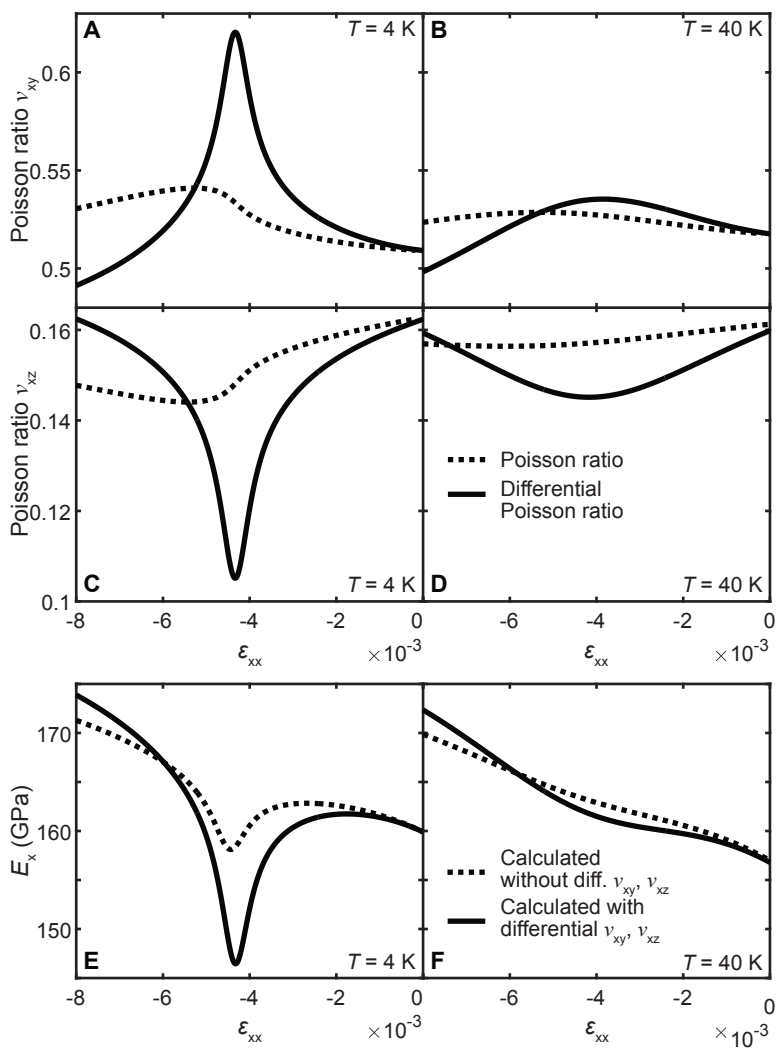

**Importance of non-linear effects at low temperatures and high strains.** (**A-D**) Standard (dotted lines) and differential (solid lines) Poisson ratios as a function of strain calculated within our model at 4 K (A, C), and at 40 K (B, D). (A, B) *xy* Poisson ratio; (C, D) *xz* Poisson ratio. (**E, F**) Young's modulus as a function of strain calculated without using differential Poisson ratios (dotted lines) and including them (solid lines) at 4 K (E) and at 40 K (F). At low temperatures, calculating $E_x$ without differential Poisson ratios underestimates the depth of the VHS minimum by over a factor of two (E).

**Fig. S7.**

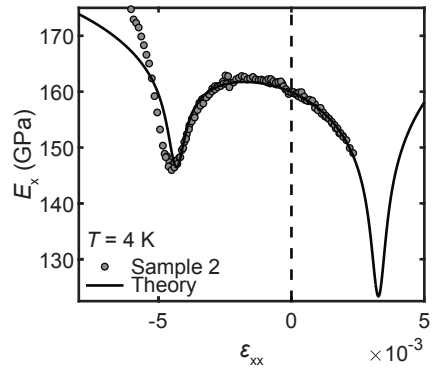

**The measured and calculated Young's modulus $E_x$ of $Sr_2RuO_4$ as a function of strain $\varepsilon_{xx}$.**
While the experimental data (grey points) stop at moderate tension because the sample broke, they follow the theory curve (black line) which crosses the tensile Lifshitz transition at $\varepsilon_{xx} = +0.33\%$. Data from Sample 2 are reproduced from Fig. 1A of the Main Text. Theory data at negative strains are also plotted in Figs. 2A, B of the Main Text.

**Fig. S8.**

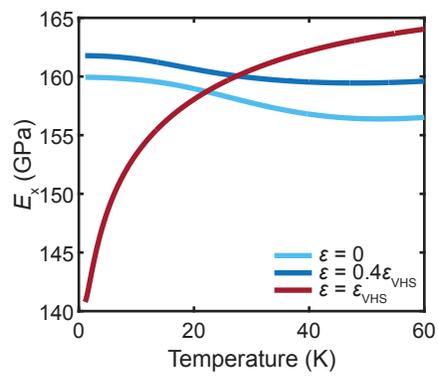

$E_x$ **calculated as a function of temperature at three key strains.** The strains chosen here are the same as those of the experimental data plotted in Fig. 3C of the Main Text. Red, $\varepsilon_{VHS}$; dark blue, $0.4\varepsilon_{VHS}$; light blue, zero strain.

**Fig. S9.**

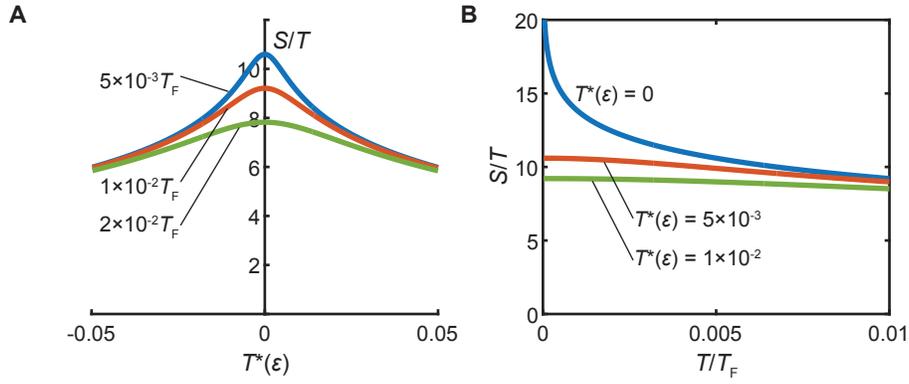

**Strain and temperature dependence of the entropy in an electronic system where the Sommerfeld coefficient diverges logarithmically at $T^*(\varepsilon) = A(\varepsilon - \varepsilon_c) = 0$.** (**A**) The entropy $S$ divided by temperature $T$ has a maximum at $T^*(\varepsilon) = 0$, and (**B**) diverges with decreasing temperature at $T^*(\varepsilon) = 0$.

**Fig. S10.**

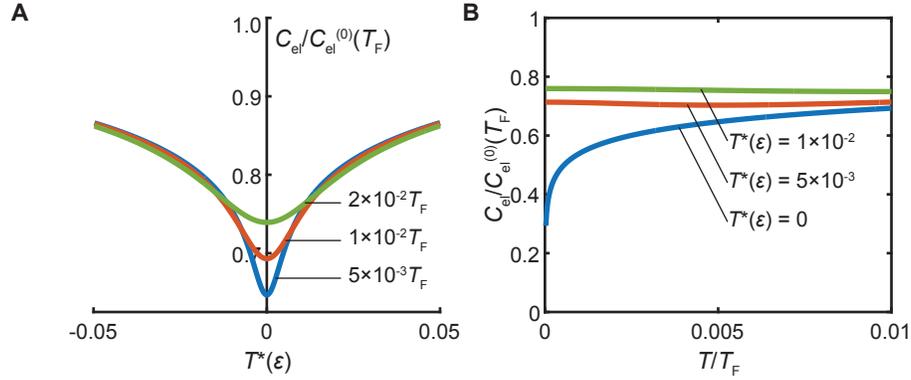

**Behavior of the elastic constant $C_{el}$, normalized by its background value $C_{el}^{(0)}(T_F)$, as a function of strain and temperature.** (**A**) $C_{el}/C_{el}^{(0)}$ vs. $T^*(\varepsilon)$ has a minimum at the critical strain $T^*(\varepsilon) = 0$ and the magnitude of the softening increases with decreasing temperature. (**B**) $C_{el}/C_{el}^{(0)}(T_F)$ vs. $T/T_F$ shows a strong temperature dependence at $T^*(\varepsilon) = 0$; away from the critical strain, $C_{el}/C_{el}^{(0)}(T_F)$ depends only very weakly on temperature.

**Fig. S11.**

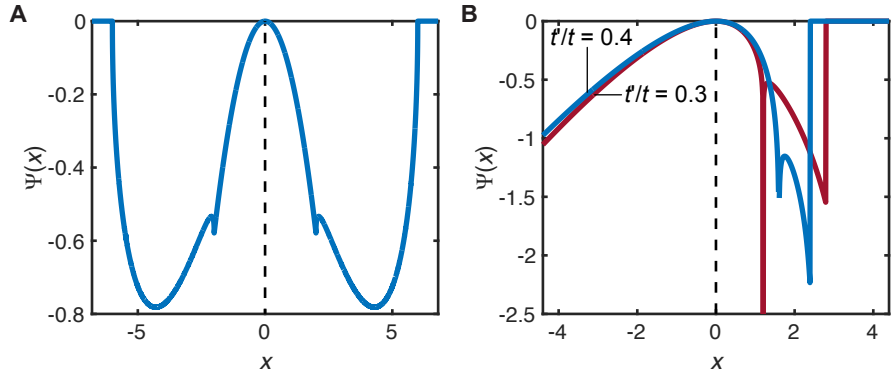

**Dimensionless function $\Psi(x)$ characterizing the contribution of conduction electrons to the elastic constant,** (**A**) for a three-dimensional system, $d = 3$, and (**B**) for a two-dimensional system, $d = 2$, with finite values of next-to-nearest neighbor hopping amplitude $t'/t = 0.3$ (red) and 0.4 (blue).

**Fig. S12.**

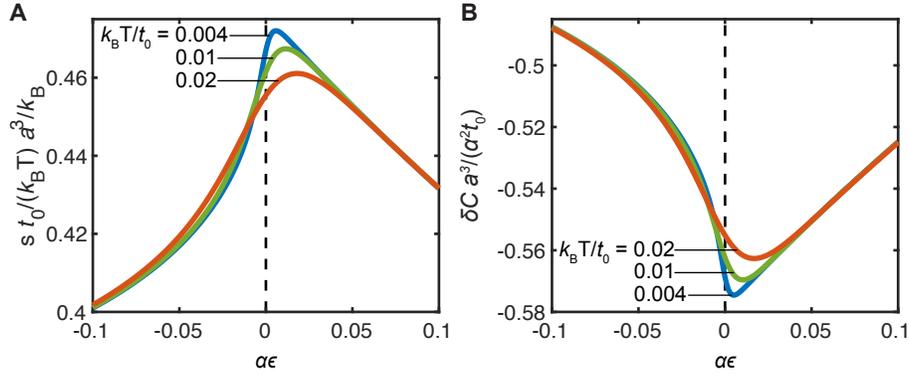

**Entropy density and elastic constant as a function of strain for a Lifshitz transition in a three-dimensional metal at various temperatures.** (**A**) Entropy density $s/T$ in units of $k_B/a^3$ divided by dimensionless temperature $k_BT/t_0$ as a function of $\alpha\epsilon$, i.e., strain, multiplied by the coefficient $\alpha$ for temperatures $k_BT/t_0 = 0.004$ (blue), 0.01 (green), and 0.02 (orange), and a chemical potential $\mu = -2t_0$. At lowest temperatures, $s/T$ is proportional to the density of states that exhibits a typical $\sqrt{\epsilon}$-cusp at the Van Hove singularity in three spatial dimensions. This cusp singularity is washed out at finite temperatures. (**B**) Correction to the elastic constant $\delta C$ in units of $t_0\alpha^2/a^3$ as a function of $\alpha\epsilon$ for temperatures $k_BT/t_0 = 0.004$ (blue), 0.01 (green), and 0.02 (orange), and a chemical potential $\mu = -2t_0$. At lowest temperatures, $\delta C$ resembles the density of states whose cusp-singularity is washed out at finite $T$.

**Fig. S13.**

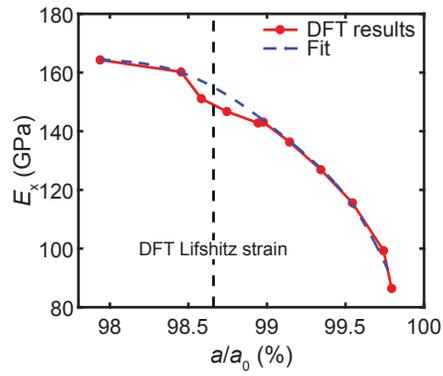

**Young's modulus of Sr$_2$RuO$_4$ along the *x*-axis, $E_x$, as a function of the normalized *a* lattice parameter, from DFT.** $a_0$ corresponds to the lattice parameter 3.880 Å and the Lifshitz stress point is at $a$ = 3.828 Å. One can clearly observe the softening of the stress start to show starting from the Lifshitz transition point. The dashed line is a fit through all points except those near the Lifshitz point.

**Table S1.**

| Sample | $L_{neck}$ (µm) | Thickness (µm) | Neck width (µm) | Anchor tab width (µm) |
|---|---|---|---|---|
| 1 | 742 | 103.6 | 123 | 376 |
| 2 | 748 | 101.8 | 93 | 801 |
| 3 | 822 | 74 | 107 | 1178 |

**Key dimensions of the stress-strain samples.** Lengths and widths were measured in images taken with a scanning electron microscope (SEM). The neck width is the average of top and bottom surfaces, measured in the central region of the neck. The thickness was measured either with an optical profiler (Samples 1 & 2) or in SEM images.

**Table S2.**

| Sample | Rig | Displacement sensor calibration | | Force sensor calibration | |
|---|---|---|---|---|---|
| | | $α_D$ (F m) | $C_{D,offset}$ (F) | $α_F$ (F N) | $C_{F,offset}$ (F) |
| 1 | Home-built #1 | 3.99e-17 | 6.23e-14 | 4.47e-10 | 6.08e-14 |
| 2 | Home-built #2 | 4.28e-17 | 6.39e-14 | 2.87e-10 | 4.02e-13 |
| 3 | UC200 | 4.46e-17 | 6.05e-14 | 3.83e-10 | 9.89e-14 |

**Calibration parameters for the capacitive displacement and force sensors in the stress-strain rigs used in this work.** The values for $α_F$ given here have been scaled to bring the Van Hove stress to -0.7 GPa.